\documentclass[iop]{emulateapj}

\usepackage{graphicx}
\usepackage{color}
\usepackage{amsmath,amssymb}
\usepackage{type1cm}
\usepackage{ulem}
\usepackage{mathrsfs}

\usepackage{xfrac}
\usepackage{hyperref}

\def\ga{\,\,\raise0.14em\hbox{$>$}\kern-0.76em\lower0.28em\hbox{$\sim$}\,\,}
\def\la{\,\,\raise0.14em\hbox{$<$}\kern-0.76em\lower0.28em\hbox{$\sim$}\,\,}

\slugcomment{draft version \today, }

\shorttitle{Global model of prompt emission in SGRBs}
\shortauthors{Ito et al.}

\begin{document}

\title{A global numerical model of the prompt emission in short gamma-ray 
bursts}

\author{Hirotaka Ito\altaffilmark{1,2},
  Oliver Just\altaffilmark{3,1},
  Yuki Takei\altaffilmark{4,5,1},
  Shigehiro Nagataki\altaffilmark{1,2}}  

\altaffiltext{1}{Astrophysical Big Bang Laboratory, RIKEN, Saitama 351-0198, Japan}
\email{hirotaka.ito@riken.jp}
\altaffiltext{2}{Interdisciplinary Theoretical \& Mathematical Science Program (iTHEMS), RIKEN, Saitama 351-0198, Japan}
\altaffiltext{3}{GSI Helmholtzzentrum f\"ur Schwerionenforschung, Planckstra{\ss}e 1, 64291 Darmstadt, Germany}
\altaffiltext{4}{Research Center for the Early Universe (RESCEU), Graduate School of Science, The University of Tokyo, 7-3-1 Hongo, Bunkyo-ku, Tokyo 113-0033, Japan}
\altaffiltext{5}{Department of Astronomy, Graduate School of Science, The 
University of Tokyo, 7-3-1 Hongo, Bunkyo-ku, Tokyo 113-0033, Japan}

\begin{abstract}
  We present the first global model of prompt emission from a short gamma-ray burst that consistently describes the evolution of the central black-hole (BH) torus system, the propagation of the jet through multi-component merger ejecta, the transition into free expansion, and the photospheric  emission from the relativistic jet. To this end, we perform a special relativistic neutrino-hydrodynamics simulation of a viscous BH-torus system,  which is formed about 500\,ms after the merger and is surrounded by dynamical ejecta as well as neutron star winds, along with a jet that is injected in the vicinity of the central BH. In a post-processing step, we compute the photospheric emission using a relativistic Monte-Carlo radiative transfer code.
  It is found that the wind from the torus leaves a strong imprint on the jet as well as on the emission causing narrow collimation and rapid time variability.   The viewing angle dependence of the emission gives rise to correlations among the spectral peak energy, $E_p$, isotropic energy, $E_{\rm iso}$, and peak luminosity $L_p$ which may provide natural explanations for the Amati- and Yonetoku-relations.      We also find that the degree of polarization is small for the emission from the jet core ($\lesssim 2 \%$), while it tends to increase with viewing angle outside of the core and can become as high as $\sim 10-40 \%$ for energies larger than the peak energy.      Finally, the comparison of our model with GRB170817A strongly disfavors the photospheric emission scenario and therefore supports alternative scenarios, such as the cocoon shock breakout.
\end{abstract}


\section{INTRODUCTION}\label{sec:introduction}

The recent gravitational wave event GW170817 \citep{AAA17} that was detected along with the gamma-ray event GRB170817A \citep[][]{GVB17, SFK17, AAA17b} has provided firm evidence that (at least some) short gamma-ray bursts (SGRBs) originate from binary neutron star (BNS) mergers.
The UV, optical and infrared kilonova counterparts (AT2017gfo) further revealed \citep[e.g.,][]{CBV17,KMB17,VGB17,Tanaka2017t} that a few percent of the solar mass of material became ejected during this event with favorable properties to enable the rapid-neutron-capture process \citep[see,][for recent reviews]{Cowan2021g,Arnould2020f}. Moreover, light curve analyses revealed that the outflow was composed of multiple ejecta components with a broad range of the electron fraction, $Y_e$ \citep[e.g.][]{KMB17,Perego2017a,KST18,Watson2019s} in agreement with previous predictions from 
theoretical models \citep[e.g.][]{Goriely2011, Korobkin2012, Fernandez2013b, Wanajo2014a, Perego2014a, Just2015a, Fujibayashi2018a, Siegel2017b}.
The late afterglow emission \citep{MNH18, MDG18, LLL19, MMB20}, and in particular the detection of superluminal motion in radio bands \citep{MDG18, GSP19, HNG19}, confirmed that a relativistic jet was indeed launched and that it successfully punched through the cloud of merger ejecta that was expelled prior to the jet formation.
Simultaneously, these observations could constrain the angular structure of the jet and found that the viewing angle (between the jet axis and the 
observer line of sight) is rather large, namely $\theta_{\rm obs} \sim 14^{\circ} - 19^{\circ}$ \citep[see e.g.,][for a review on the multi-wavelength observations]{MC20}.

The propagation of a jet through and its breakout from merger ejecta has been the subject of a large number of studies, most of which employing hydrodynamical (HD) and magnetohydrodynamical (MHD) simulations \citep[e.g.,][]{NHS14, MMR14,  DQM15,  JOJ16, MRM17,  DQK18,  GLN19,  HKI20, HI21, GNB21, PCV21}. After the discovery of GRB170817A/GW170817 many efforts were devoted to relating results from hydrodynamical simulations to properties of the emitted radiation in order to explore the range of conditions under which the observed peculiar features of the prompt gamma-ray and/or afterglow can or cannot result \citep[e.g.,][]{LLC17, GNP18, BTG18, NGP18, LPM18, XZM18, NGP20, NGP21, UDM20}. Recent studies \citep{NBR21,KDK21} also explore the impact of the jet-ejecta interaction on the kilonova emission. To our knowledge, however, no study so far has directly utilized the results of hydrodynamical simulations to compute the spectral properties of the prompt emission of ordinary SGRBs.
Such a step is, however, highly desirable in order to establish a firm link between hydrodynamical central engine models and the GRB emission. 

The origin of the GRB prompt emission has been extensively discussed over the past few decades without reaching a consensus. One of the central questions is whether the radiation is produced under optically thin conditions (``synchrotron model'') or optically thick conditions (``photospheric emission model'') \citep[see e.g.,][and references therein]{M19}. While in the former scenario dissipation takes place above the photosphere, giving rise to non-thermal synchrotron photons that freely escape from the emission site,
in the latter scenario photons decoupling from the photosphere are considered as the origin of the prompt GRB. 
  Regarding the link between jet hydrodynamics and the emission model, it is far more difficult to make robust predictions for the synchrotron model, because the involved processes are non-thermal and therefore very challenging to include in global simulations. 
  This is not the case in the photospheric emission model, where the emission is quasi-thermal, and, hence, reasonable estimates of the emission properties based on global hydrodynamical simulations can be made more readily\footnote{However, non-thermal effects that are possibly relevant in the photospheric emission scenario (such as sub-photospheric dissipation, see e.g. \citealp{VB16}) are typically not resolved by global hydrodynamical simulations.}.

Indeed, in the context of long GRBs (LGRBs) numerous attempts have already been made starting more than a decade ago to evaluate, based 
on the photospheric emission paradigm, basic features of the prompt GRB using data from hydrodynamical simulations \citep[e.g.,][]{LMB09, NIK11, MN11, LMM13, LML14}.
%
Subsequent studies employed sophisticated radiative transfer schemes that 
improved the accuracy of the predictions \citep{CAM15, IMN15, L16, PL18, PLL18, IMN19}. These studies have shown that a jet that successfully breaks out from the envelope of a massive star can produce photospheric emission with overall properties (such as luminosity and spectral peak energy) that are compatible with typical observed GRBs.
Moreover, it was found \citep[e.g.,][]{PL18, IMN19} that the angular structure of the released jet, which develops mainly via the interaction with the stellar envelope, gives rise to a correlation between the luminosity and peak energy for different viewing angles, which is broadly consistent 
with the empirical Amati- and Yonetoku-relations that are found to hold for observed GRBs \citep{AFT02, YMN04}.
 To gain further insight, \citet{PLL20} has recently updated their transfer scheme to explore also the polarization signature imprinted in the photospheric emission.
Although radiative transfer was not implemented, the recent series of papers by \citet{GLN19, GLN20, GBS20, GBL21} examined the radiation efficiency of photospheric emission based on high-resolution 3D HD and MHD simulations under various conditions.
%



In this paper, we simulate the photospheric emission radiated from an SGRB 
jet by applying a Monte-Carlo solver that was developed and applied in previous studies of LGRBs \citep{IMN15, IMN19}. To this end, we utilize the 
results from a hydrodynamical simulation of a relativistic jet that breaks out from BNS merger ejecta and, in a post-processing step, compute the 
radiative transfer of $\gamma$ photons on the hydrodynamic background.
We assume the jet to be launched from a BH-torus system that is formed after the metastable NS (MNS) remnant, which results immediately after the merger, collapses to a BH. One notable feature of our hydrodynamic model is that it includes the evolution of the central BH-torus system. Therefore, our model takes into account not only the interaction of the jet with 
the dynamical ejecta and the MNS winds, as in the case of most previous HD simulations, but also the effects of the jet-torus interaction.
As for the radiative transfer calculation, we employ an upgraded version of our Monte-Carlo solver that is now capable of extracting the polarization signature of the emitted radiation.



The primary goal of this first study solving photospheric radiative transfer on a hydrodynamic background of an SGRB simulation is to get a basic idea of the light curve, spectra, and polarization properties and to roughly assess in which aspects our model does and does not agree with observations. While we will also briefly compare with the peculiar GRB170187A, the main focus here is on ordinary SGRBs.

This paper is organized as follows: In \S\ref{sec:model}, we describe our 
hydrodynamical model and the employed radiative transfer methods. The main results will be presented in \S\ref{sec:result}, while implications obtained from the comparison with GRB170817A are discussed in \S\ref{sec:1701817}.
\S\ref{sec:summary} provides the summary and discussion.

\section{MODEL AND METHODS}
\label{sec:model}



\subsection{Hydrodynamical simulation}\label{sec:hydr-simul-1}

The hydrodynamical model is obtained using the ALCAR code \citep{Obergaulinger2008a, JOJ15}, which employs spherical coordinates and Riemann-solver based finite-volume methods to solve the special relativistic hydrodynamics equations in 2D axisymmetry coupled to two-moment neutrino transport 
that is described by a multi-energy group M1 scheme. The numerical setup is similar as in \citet{JOJ15, JOJ16, Just2021a}. As described in detail below, the main differences to the models of \citet{JOJ16} are a parameterized description of the merger ejecta surrounding the torus (instead of mapping the ejecta from a previously performed merger simulation) and the fact that we now inject a jet manually with predefined properties (instead of following the jet that could eventually be launched due to heating from neutrino-pair annihilation).

We describe turbulent angular momentum transport using the $\alpha$-viscosity approach \citep{Shakura1973}, where the dynamic viscosity is computed as $\eta=0.03 \rho c_s^2/\Omega_{K}$ (with baryonic mass density $\rho$, isothermal sound speed $c_s^2=P/\rho$, gas pressure $P$, and Keplerian angular velocity $\Omega_{K}$). We assume that the central BH has a mass of $M_{\mathrm{BH}}=2.7\,M_\odot$ and a spin parameter of $A_{\mathrm{BH}}=0.8$, and we employ a pseudo-Newtonian gravitational potential \citep{Artemova1996} to approximately include basic general relativistic effects.

\begin{figure*}[htpb]
\begin{center}
\includegraphics[trim=0 15 0 15,clip,width=0.8\textwidth,keepaspectratio]{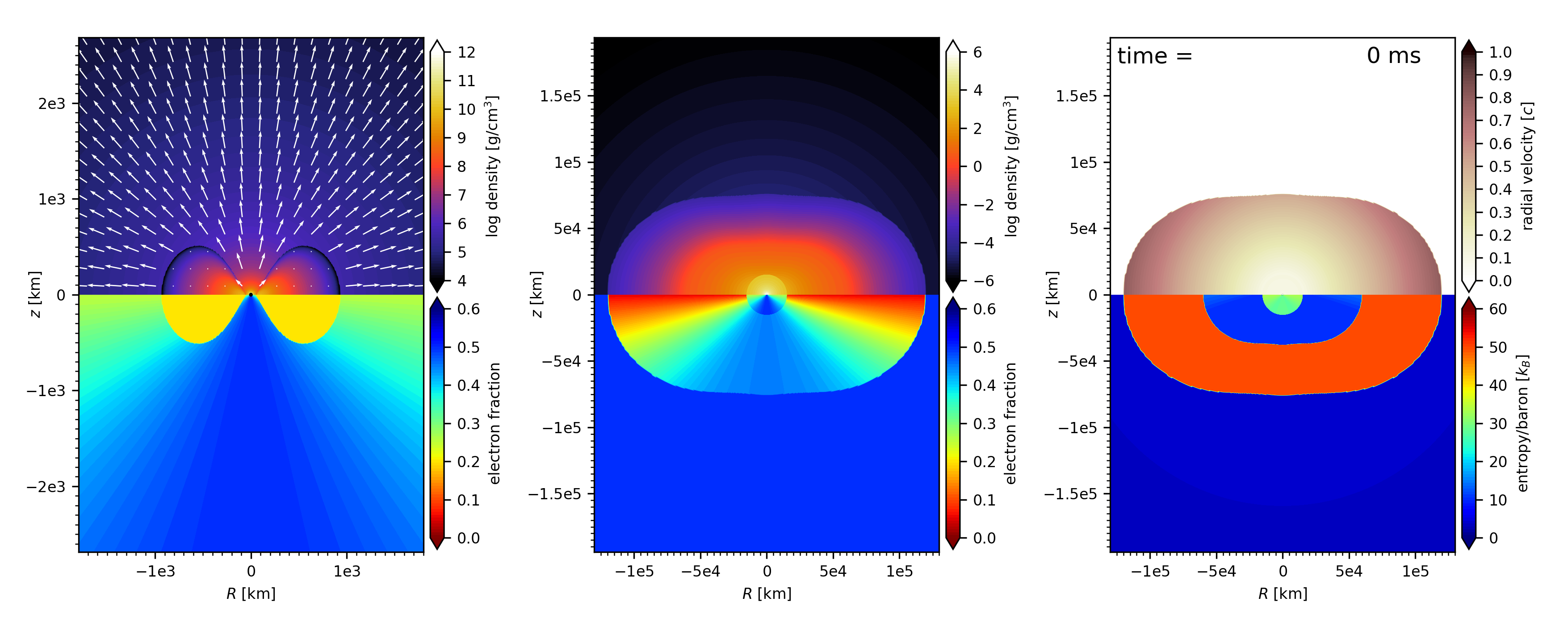}
\includegraphics[trim=0 15 0 15,clip,width=0.8\textwidth,keepaspectratio]{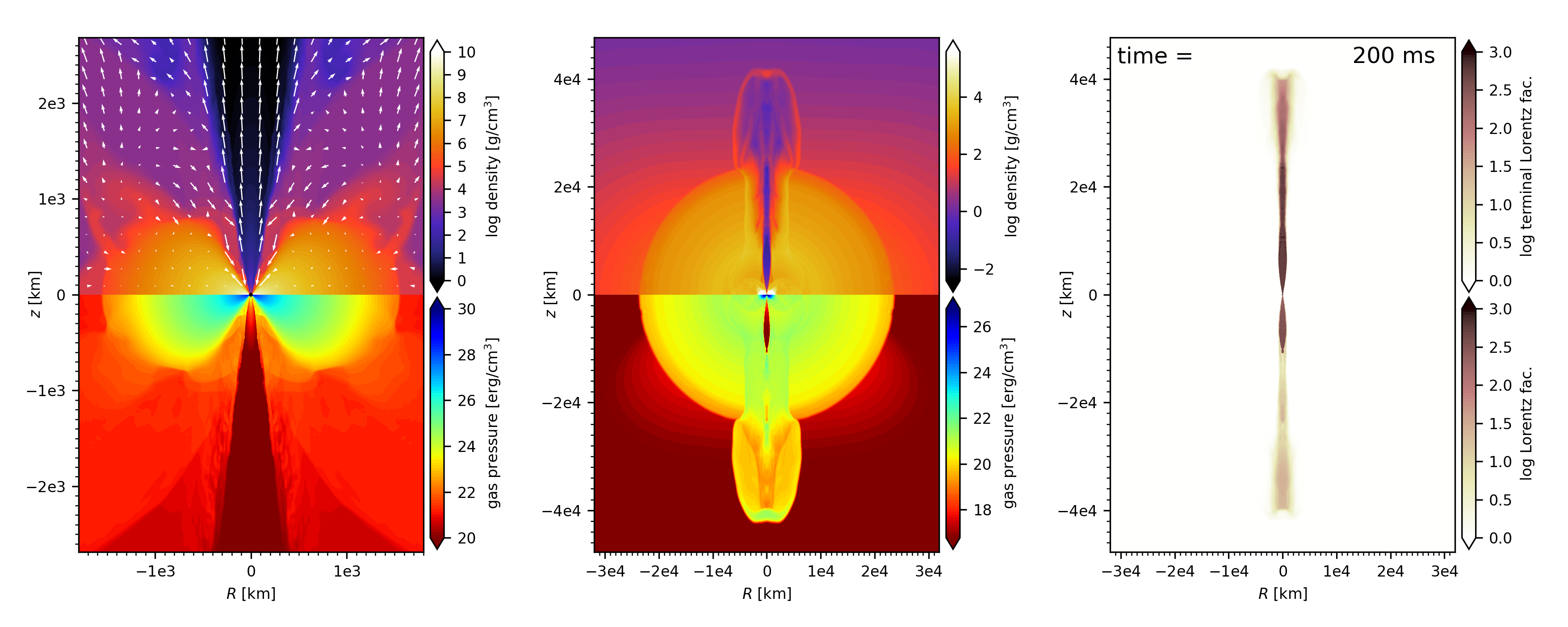}
\includegraphics[trim=0 15 0 15,clip,width=0.8\textwidth,keepaspectratio]{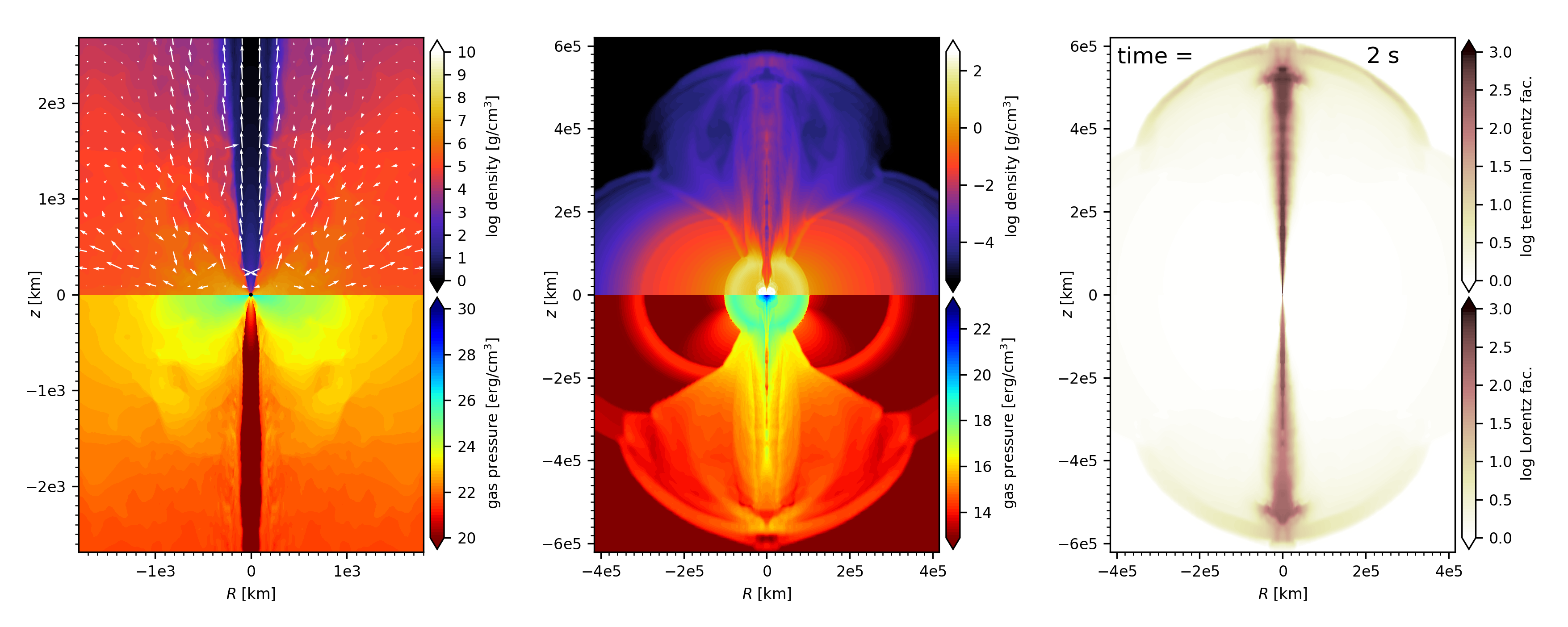}
\includegraphics[trim=0 15 0 15,clip,width=0.8\textwidth,keepaspectratio]{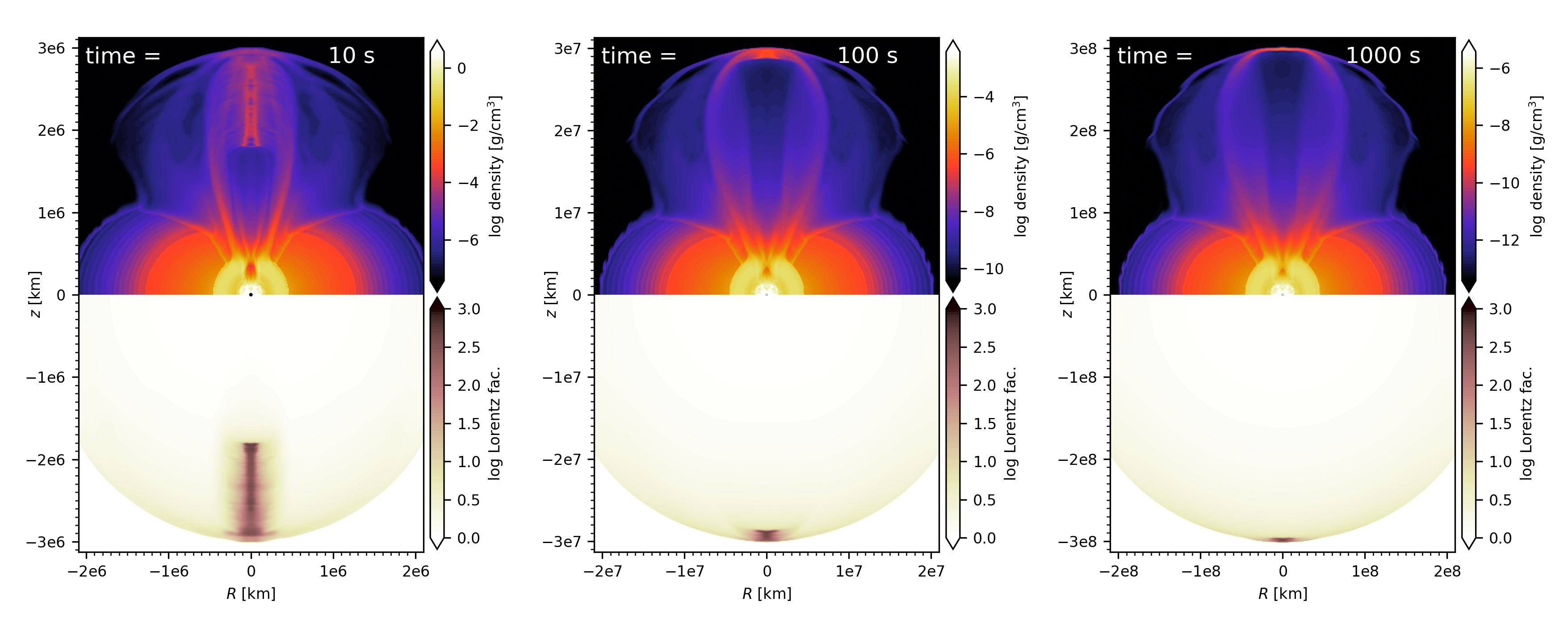}
\end{center}
\caption{Snapshots illustrating different phases of evolution of our hydrodynamical model. {\it Top row:} Initial configuration at 500\,ms post merger. At this time the radial extent of the torus, MNS wind, core dynamical ejecta, and fast tail dynamical ejecta are $10^3\,$km, $1.5\times 10^4\,$km, $6\times 10^4\,$km, and $1.2\times 10^5\,$km, respectively, along the equatorial plane. {\it Second row:} After jet breakout from the MNS ejecta. {\it Third row:} After jet breakout from the dynamical ejecta. {\it Bottom row:} Free expansion and transition into optically thin conditions at times $t=10\,$s, 100\,s, 
and 1000\,s from left to right. Color maps show the quantities provided on the right side of each panel, while white arrows indicate velocity vectors.}
\label{fig:hydrocontours}
\end{figure*}

The initial configuration (i.e. at $t=0$, see top panel of Fig.~\ref{fig:hydrocontours} for contour maps) consists of a BH-torus system that is assumed to be formed $\tau_{\mathrm{MNS}}=500$\,ms after\footnote{The time of collapse, $\tau_{\mathrm{MNS}}$, is, for given BNS parameters, not 
well constrained, because it is sensitive to the unknown nuclear equation 
of state and to yet poorly understood processes responsible for transporting energy and angular momentum out of the MNS \citep[e.g.][]{Shibata2005d, Duez2006a, Paschalidis2012l, Kiuchi2014, Kastaun2015a, Radice2018f, Ciolfi2019a}. As for GW170817, values of $\tau_{\mathrm{MNS}}\ga 10\,$ms are required to explain the high luminosity of the blue kilonova, while $\tau_{\mathrm{MNS}}\la 1-2\,$s is needed to explain GRB170817A with a BH-torus central engine \citep[e.g.][and references therein]{Metzger2017d}.} a 
BNS merger following the collapse of the metastable NS remnant (MNS). We model the torus, the dynamical ejecta, as well as the wind ejecta launched during the lifetime of the MNS (called MNS ejecta in the following) with properties that are guided by results of previous merger and post-merger simulations \citep[e.g.][]{Bauswein2013, Foucart2016a, Sekiguchi2015a, Fujibayashi2018a, Radice2018b}. The torus is prescribed as an equilibrium 
configuration with a given mass (0.044\,$M_\odot$), constant entropy per baryon ($10\,k_B$), given radius of maximum density ($100\,$km), and a cylindrical rotation profile (with angular velocity $\propto(r\sin\theta)^{-1.8}$). The torus is surrounded by MNS ejecta reaching from the radial inner boundary out to $r=0.1\,c\,\tau_{\mathrm{MNS}}$ (with $c$ being the speed of light). This ejecta component is modeled as a spherically symmetric wind with a given mass flux ($\approx 0.07\,M_\odot\,$s$^{-1}$) and 
entropy per baryon ($30\,k_B$). Finally, exterior to the MNS ejecta we place the dynamical ejecta, which is assumed to be launched at the time of 
the merger ($t=-\tau_{\mathrm{MNS}}$). They are represented by a homologously expanding (i.e. with a radial velocity of $v(r,t=0)=r/\tau_{\rm MNS}$) gas cloud of mass $5\times 10^{-3}\,M_\odot$ with a continuous density profile following $\rho(v,\theta)\propto v^n(0.25+\sin^3\theta)$, where $n=-3.5$ for $v\in[0.1c,0.4c]$ and $n=-14$ for $v\in[0.4\,c,0.8\,c]$ (inspired by a similar profile used in \citealp{Gottlieb2018a}). We 
note that most existing simulations of jets from merger remnants only include a single, homologously expanding gas cloud, whereas our setup explicitly accounts for two ejecta components. Compared to the dynamical ejecta, the MNS ejecta exhibit a more shallow density profile (roughly $\rho\propto r^{-2}$). Furthermore, they are not homologous and carry initially a 
significant amount of thermal energy. Finally, we set the initial electron fraction, $Y_e$, as follows: The torus has a constant $Y_e$ of $0.2$, in the MNS ejecta $Y_e$ increases linearly with $\cos\theta$ from 0.25 (at 
the equator) to 0.5 (at the pole), and in the dynamical ejecta $Y_e$ is monotonically increasing from 0.05 (at the equator) to 0.45 (at the pole) in such a way that the differential mass distribution $\mathrm{d}m/\mathrm{d}Y_e\approx \mathrm{const.}$.

As our special relativistic simulation is unable to describe the general relativistic Blandford-Znajek process  \citep{Blandford1977, TN08}, and since neutrino annihilation is not efficient enough to create a funnel and launch 
a powerful jet \citep{JOJ16}, we inject the jet in a parametric way at 
the inner radial boundary. The properties of the injected jet are determined by fixing its half-opening angle ($\theta_{j,\rm ini} = 10^{\circ}$), power ($L_j = 10^{50}$\,erg\,s$^{-1}$), Lorentz factor ($\Gamma_{j,\rm ini} = 5$), specific enthalpy ($h_{\rm ini} = 100$), and electron fraction ($0.5$). The jet injection starts at $t=10\,$ms and stops at $t=4\,$s. The total energy of the jet injected into one hemisphere is thus $\approx 4\times10^{50}\,$erg and its terminal Lorentz factor is $\Gamma_{j,\rm ini} h_{\rm ini}=500$. 

In order to be able to follow the photospheric emission for a sufficiently wide range of viewing angles, we need to evolve the jet up to radii of $r\sim 10^{14}\,$cm \citep{IMN19}\footnote{Within the range of viewing angles considered in this study ($\theta_{\mathrm{obs}}\leq 4.5^{\circ}$), the largest average radius of last scattering at a given time is found to lie at about $4 \times 10^{13}~{\rm cm}$.}. To this end, we first evolve the central engine until about the end of its activity ($t\approx 4\,$s) on a relatively small domain ranging from $r_{\mathrm{in}} = 10^6~{\rm cm}$ to $r_{\mathrm{out}} \approx 2.7\times 10^{11}~{\rm cm}$ that is sampled by 1600 logarithmically distributed grid cells. We subsequently cut out the innermost part and extend the grid in radial direction. That is, we map the configuration onto a new radial grid with boundaries $r_{\mathrm{in}}$ 
and $r_{\mathrm{out}}$ lying at $10^9\,$cm and $4\times 10^{12}\,$cm, respectively, and consisting of 25600 grid cells, which grow by a constant factor of $0.016\,\%$ from cell to cell. We repeat this remapping step twice at evolution times of $100\,$s and $1000\,$s, each time using 10 times 
larger values for $r_{\mathrm{in/out}}$ but otherwise keeping the same grid properties. The angular grid covers only the northern hemisphere (i.e. 
we assume equatorial symmetry) and consists of 160 cells. While at late times this grid is uniform in $\theta$, during the first $4\,$s of evolution, namely when the jet travels through the ejecta and breaks out, we employ a non-uniform distribution with slightly enhanced resolution in the polar region by using $\Delta \theta\approx 0.28^\circ$ for $\theta\la 23^\circ$ and $\Delta \theta\approx 0.95^\circ$ for $\theta\ga 23^\circ$ with smooth transition of $\Delta \theta$ in between.

Our grid-based simulation code (as well as many other ones) does not evolve the thermal energy density directly, but instead the sum of the thermal, kinetic, and (a fraction of) rest-mass energy density. As a consequence, the thermal energy is beset with potentially large numerical truncation errors, especially in the late phase of expansion once the thermal energy density becomes smaller than about $0.1-1\,\%$ of the kinetic energy. Although at this late stage of evolution the thermal energy has only a minor relevance for the dynamics of the flow, it determines the temperature 
and therefore the photon content of the radiating jet. Hence, we wish to track it as accurately as possible. For that reason we additionally evolve the conservation equation for the entropy, i.e. $\partial_t (\Gamma \rho s) + \nabla_j(\Gamma \rho v^j s) = 0$ (with Lorentz factor $\Gamma$, entropy per baryon $s$ and fluid velocity three-vector $v^j$). We use the above equation to compute the thermal energy and the temperature, but only for radii $r>10^3\,$km and not in the vicinity of shocks. In all other cases, we use the conventional energy equation. We keep both equations synchronized with each other by resetting 
the corresponding other thermodynamic quantity after each time step. Our treatment is motivated by a similar procedure implemented in the PLUTO code \citep{Mignone2007a}.

\subsection{Radiative transfer calculation}\label{sec:radi-transf-calc}

The results of the hydrodynamical simulation serve as input for the subsequent radiative transfer calculation. To this end, we employ the Monte-Carlo code that was already used in \citet{IMN15, IMN19} for studying LGRBs. 
We refer to these papers for a description of the technical details.
Since the neutrino-hydrodynamical simulation takes into account weak interactions (cf. Sect.~\ref{sec:hydr-simul-1}), we take the comoving number density of the electrons, $n_e$, which determines the photon opacity, to be consistent with the evolved electron fraction $Y_e$, i.e. $n_e = Y_e 
\rho / m_p$ (where $m_p$ is the proton rest mass). While this treatment differs from the one in \citet{IMN15, IMN19}, in which $Y_e = 1$ was assumed, the impact is small, because the electron fraction in the jet remains close to $Y_e \sim 0.5$ at all times.
Moreover, compared to the aforementioned studies, we now track not only the energy-momentum flow of the photons but also their polarization state. 
For that purpose, we implemented the method described in \citet{INM14}, which we briefly outline in the following.

The Monte-Carlo radiative transfer simulation tracks the evolution of photon packets sampling the radiation field, which are initially injected in 
regions of high optical depth.
Each photon packet is characterized by the photon frequency $\nu$ as well 
as the Stokes parameters, $I$, $Q$, and $U$. Here, $I=n_{\rm pack} h\nu$ 
is the intensity of the electromagnetic wave, which is taken to be equal to the total energy carried by the packet, where $n_{\rm pack}$ is the number of photons per packet and $h$ is the Planck constant. Each photon packet contains the same number of photons. The quantities $Q$ and $U$ measure the linear polarization. We do not take into account the Stokes parameter $V$ characterizing circular polarization, because electron scattering
does not induce circular polarization as long as the spin of electrons has isotropic distribution.
%
We initially inject about $7.5 \times 10^8$ photon packets in total in a way to sample the local Planck distribution corresponding to the local temperature provided by the hydrodynamical simulation, assuming a vanishing 
degree of polarization ($Q=U=0$). After the injection, we track the evolution of each packet under the influence of scatterings until it reaches the outer boundary that is located well above the photosphere. For the scattering reactions, we take into account the full Klein-Nishina cross-section and self-consistently calculate the change in the Stokes parameters. The spectrum and polarization are finally evaluated by binning the packets at the boundary. Regarding the coordinate system employed to define $Q$ and $U$, we use the same prescription as in \citet{INM14} (see their Fig.~4). Hence, a positive (negative) value of $Q$ corresponds to the case when the electric vector of the polarized beam is parallel (perpendicular) to the plane formed by the line of sight (LOS) of the observer and the jet axis. Due to the assumption of axisymmetry in the current simulation, the resulting polarization is always aligned to either of the two aforementioned directions and the parameter $U$ summed over all packets must 
vanish exactly in the analytical limit. We confirmed that our numerical results are consistent with this constraint, i.e. $|U| \ll |Q|$ everywhere.
In summary, the polarization state is fully characterized by the quantity 
$Q/I$, of which the absolute value measures the degree of polarization and its sign defines the angle of the linear polarization.

Throughout the paper, the location of the observer is expressed by the viewing angle $\theta_{\rm obs}$, which is the angle between the LOS  and the central axis of the jet.

\section{RESULTS}
\label{sec:result}

\subsection{Hydrodynamical simulation}\label{sec:hydr-simul}

\begin{figure*}[htpb]
\includegraphics[width=0.95\textwidth,keepaspectratio]{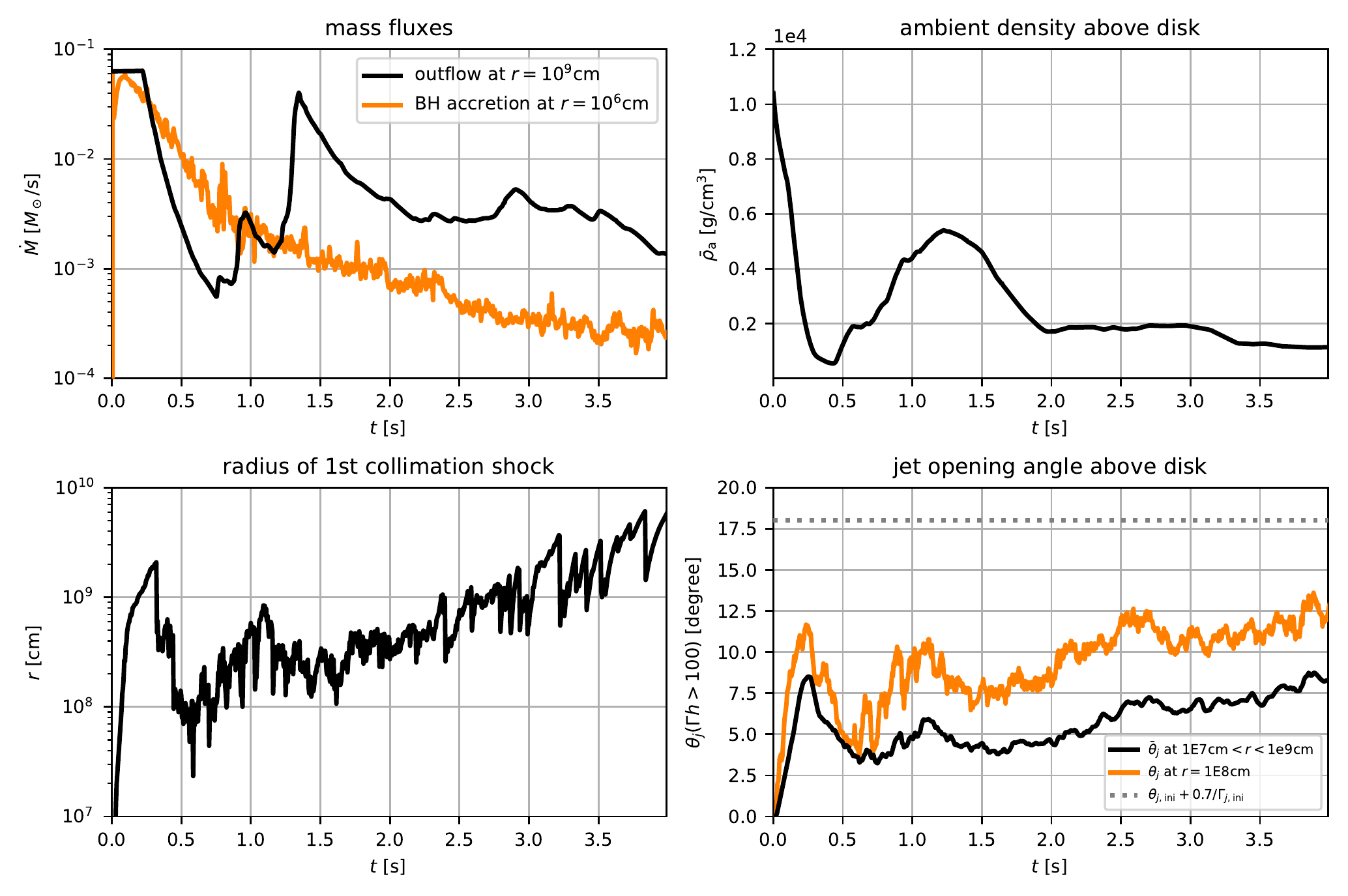}
\caption{Global properties of the hydrodynamical model. The time coordinate refers to the simulation, i.e. $t=0$ corresponds to 500\,ms post merger. {\it Top left:} Mass fluxes into the BH (orange line) and out of the sphere at $r=10^9\,$cm (black line). {\it Top right:} Average density of the jet-confining material located in a cone within $10^7\,\mathrm{cm}<r<10^9\,$cm and $\theta_{j,\rm ini}<\theta<2\,\theta_{j,\rm ini}$ around the jet. {\it Bottom left:} Radius of the tip of the first collimation shock at $\theta=0$. {\it Bottom right:} Opening angle of the ultrarelativistic region where the terminal Lorentz factors $\Gamma h>100$ measured in the vicinity of the BH-torus system, namely at $r=10^8\,$cm (orange line) and averaged over $10^7\,\mathrm{cm}<r<10^9\,\mathrm{cm}$ (black line, computed as $\bar\theta_j=\int_{10^7\,\mathrm{km}}^{10^9\,\mathrm{km}} \theta_j(\Gamma h>100)\mathrm{d}r/(0.99\times 10^9\,\mathrm{km})$). The grey dotted line corresponds to the opening angle $\theta_j\approx\theta_{j,\rm ini}+0.7/\Gamma_{j,\rm ini}$ \citep[e.g.][]{Harrison2018o} that would approximately result without any external medium collimating the jet. The disk winds launched after $t\ga 0.3-1\,$s collimate the jet and feed it with perturbations that originate from disk turbulence.\label{fig:timeplots_hydro}} 
\end{figure*}

\begin{figure}[htpb]
\begin{center}
  \includegraphics[width=9cm,keepaspectratio]{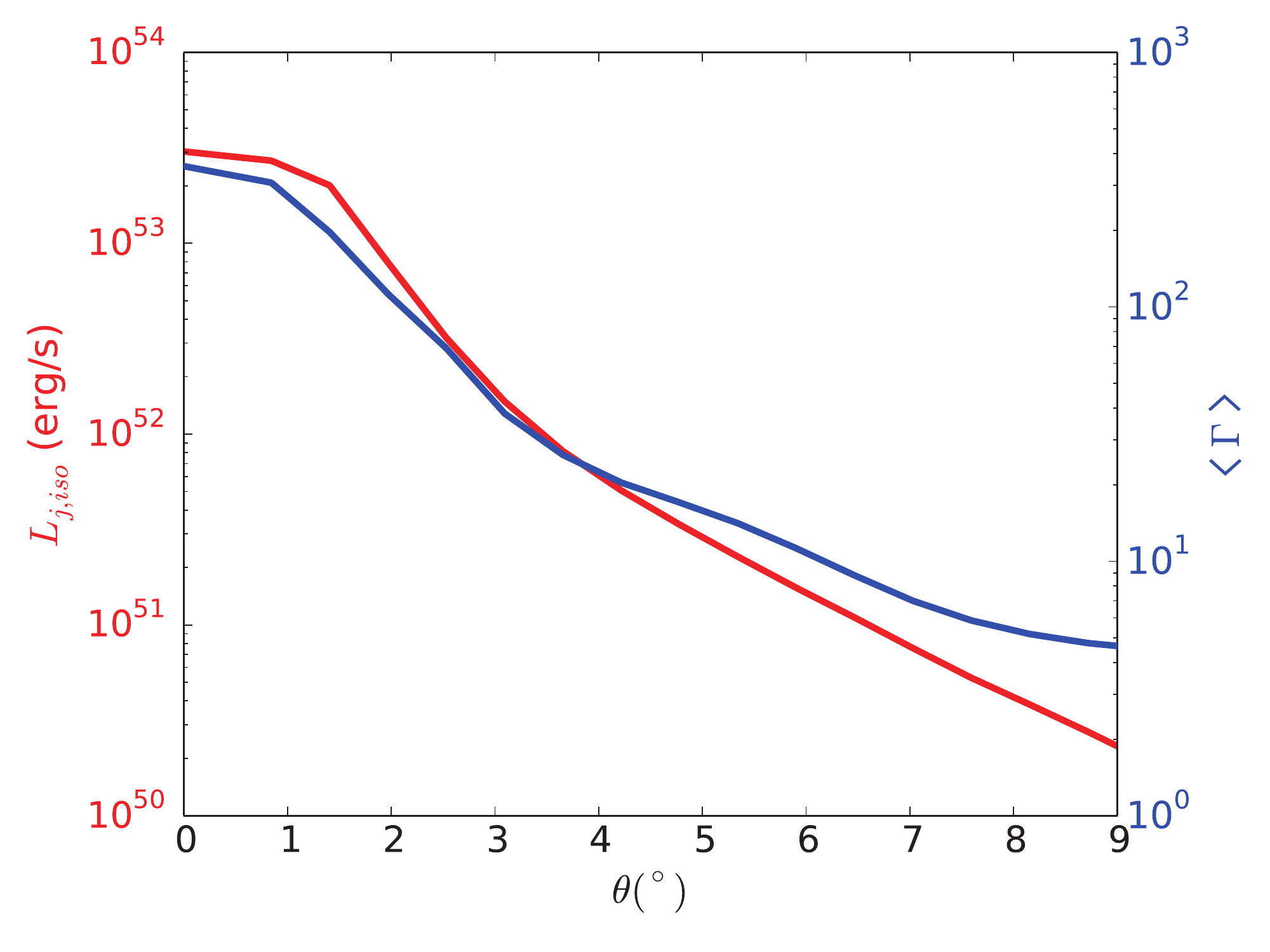}
\end{center}
\caption{Angular distribution of the time averaged istropic equivalent power $\left< L_{j,\rm iso} \right>$ ({\it red}) and Lorentz factor $\left< 
\Gamma \right>$ ({\it blue}) of the outflow evaluated at $r=10^{12}~{\rm cm}$. Both quantities are evaluated at $r=10^{12}\,$cm by integrating 
over a duration of $5~{\rm s}$ starting when the jet first reaches this radius. The Lorentz factor is computed as an energy weighted average.\label{fig:3}}
\end{figure}

In Fig.~\ref{fig:hydrocontours} we show snapshots of the hydrodynamical model at different times. As found in previous jet simulations \citep[][]{NHS14, MMR14,  DQM15, MRM17, LLC17, DQK18, GNP18,  GLN19,  HKI20, HI21, GNB21}, the interaction between the jet and the merger ejecta results in the formation of a cocoon, which collimates the jet and shapes its structure. However, different from these studies, in which a single ejecta component was considered, the jet in the current simulation experiences two breakouts, namely first from the MNS wind (at $t\sim 120~{\rm ms}$) and then from the dynamical ejecta with an extended fast tail (during $t\sim 350 
- 700~{\rm ms}$). Since the dynamical ejecta is significantly less massive than the MNS ejecta, we observe that the jet becomes wider (i.e. less collimated) once it breaks out of the MNS ejecta and travels through the dynamical ejecta.

In Fig.~\ref{fig:timeplots_hydro} we provide plots of several global quantities as functions of time. While the entire dynamical ejecta is gravitationally unbound at $t=0$, a small fraction of the MNS ejecta close to 
the central BH remains gravitationally bound and falls into the central BH within the first milliseconds. The unbound part of the MNS ejecta keeps 
expanding and crosses the sphere at $r=10^9\,$cm with a mass flux rate that is roughly constant until $t\approx 200\,$ms and afterwards drops quickly (cf. top left panel of Fig.~\ref{fig:timeplots_hydro}).

The evolution of the neutrino-cooled, viscous torus that is embedded in the previously ejected material is similar to the evolution without the ambient ejecta, which was extensively studied in earlier works \citep{Fernandez2013b, JOJ15, Siegel2017b, Fujibayashi2020a, Just2021a}. The aspect that is particularly relevant to the jet propagation is that the disk releases a powerful wind on a timescale of seconds, which is launched mainly as a result of viscous heating once neutrino cooling becomes inefficient after a few hundred milliseconds of evolution\footnote{Winds from neutrino-cooled disks are believed to be a prolific site for the production of heavy r-process elements and in the case of GW170817 may have powered 
the red kilonova component \citep[e.g.][]{Just2015a, KMB17, Siegel2017b, FTQ19}.}. Measured at $r=10^9\,$cm, the mass flux connected to 
this disk outflow peaks at about $t\approx 1.3\,$s and then gradually decreases (cf. top left panel of Fig.~\ref{fig:timeplots_hydro}). As a result, the volume surrounding the base of the jet, out of which MNS ejecta escape until about $t\sim 200\,$ms, gets replenished with disk wind material. This can be seen by the extended bump in the average density, $\bar\rho_a$ (cf. top right panel of Fig.~\ref{fig:timeplots_hydro}), of the ambient medium through which the jet propagates. We estimate this quantity as 
$\bar\rho_a\approx\int_{\mathcal C}\rho \mathrm{d}V /\int_{\mathcal C}  \mathrm{d}V$, where $\mathcal C$ denotes the cone that is defined by $10^7\,$cm$<r<10^9$\,cm and $\theta_{j,\rm ini}<\theta<2\theta_{j,\rm ini}$.

Consistent with previous simulations that resolve the jet-torus interaction \citep{AJM05, JOJ16, CLT19}, the massive torus winds efficiently confine the jet. The disk thus provides an additional collimating agent next to the other ejecta components surrounding the disk. In doing so, it basically determines the ``initial'' jet opening angle that is commonly used as a free parameter in studies that ignore the evolution of the central engine and inject the jet directly into the (dynamical or MNS) ejecta. The impact of the disk wind on the jet opening angle, $\theta_j$, and on the maximum radius of the first collimation shock \citep[e.g.][]{Komissarov1997, Bromberg2011} can be seen in the bottom panels of Fig.~\ref{fig:timeplots_hydro}: At early times, before substantial disk winds are encountered ($t\la 300-500\,ms$), the jet is rather wide, with opening angles $\theta_j\approx 10^\circ$, and the collimation shock reaches out to $\sim 10^9\,$cm. Subsequently, as massive winds are expelled by the disk, the jet becomes more narrow, with $\theta_j\approx 4^\circ$, and the collimation shock shrinks to radii of about $r\sim 10^8\,$cm. Then, after about $\sim 
1\,$s and roughly on a timescale of seconds, the jet becomes gradually wider again and the collimation shock expands, because the disk mass, and therefore the baryon loading of the wind, continuously decreases. Apart from collimating the jet, the disk winds, which exhibit time-dependent mass 
fluxes induced by the turbulent disk, also imprint significant radial and 
temporal variations on the jet. This is indicated by the non-steady behavior of the opening angle and the radius of the collimation shock (see Fig.~\ref{fig:timeplots_hydro}).

Considering the angular distribution of the jet in the free expansion phase (i.e. well after the breakout from the ejecta; see Fig.~\ref{fig:3}), we observe, as commonly found in previous studies,
a relativistic ``core'', where the jet power and Lorentz factor are high and close to uniform, and an extended ``wing'' region originating from the jet-cocoon interface \citep[e.g.][]{GNB21}, in which these quantities rapidly decrease with polar angle. Remarkably, the opening angle of the core\footnote{Following \citet{GNB21}, we define $\theta_{\rm core}$ as the 
angle in which the isotropic equivalent energy (or, equivalently, the isotropic equivalent jet power, $\left<L_{j, \mathrm{iso}}\right>$) drops to 
$75\,\%$ of its value at the axis.}, $\theta_{\rm core} \sim 1.2^{\circ}$, turns out to be rather small compared to simulations neglecting the torus evolution and using a similar initial opening angle at the injection point \citep[e.g.,][]{GBL21}. The narrow core is most likely a result of the aforementioned jet-torus interaction, which collimates the jet to opening angles $\theta_j<\theta_{j, \rm ini}$ already at an early stage. However, a closer analysis of the conditions that determine the final ejecta properties is out of the scope of the present study and is deferred to future work.

\subsection{Light curves and spectra}\label{sec:light-curves-spectra}

\begin{figure*}[htbp]
\begin{center}
\includegraphics[width=17cm,keepaspectratio]{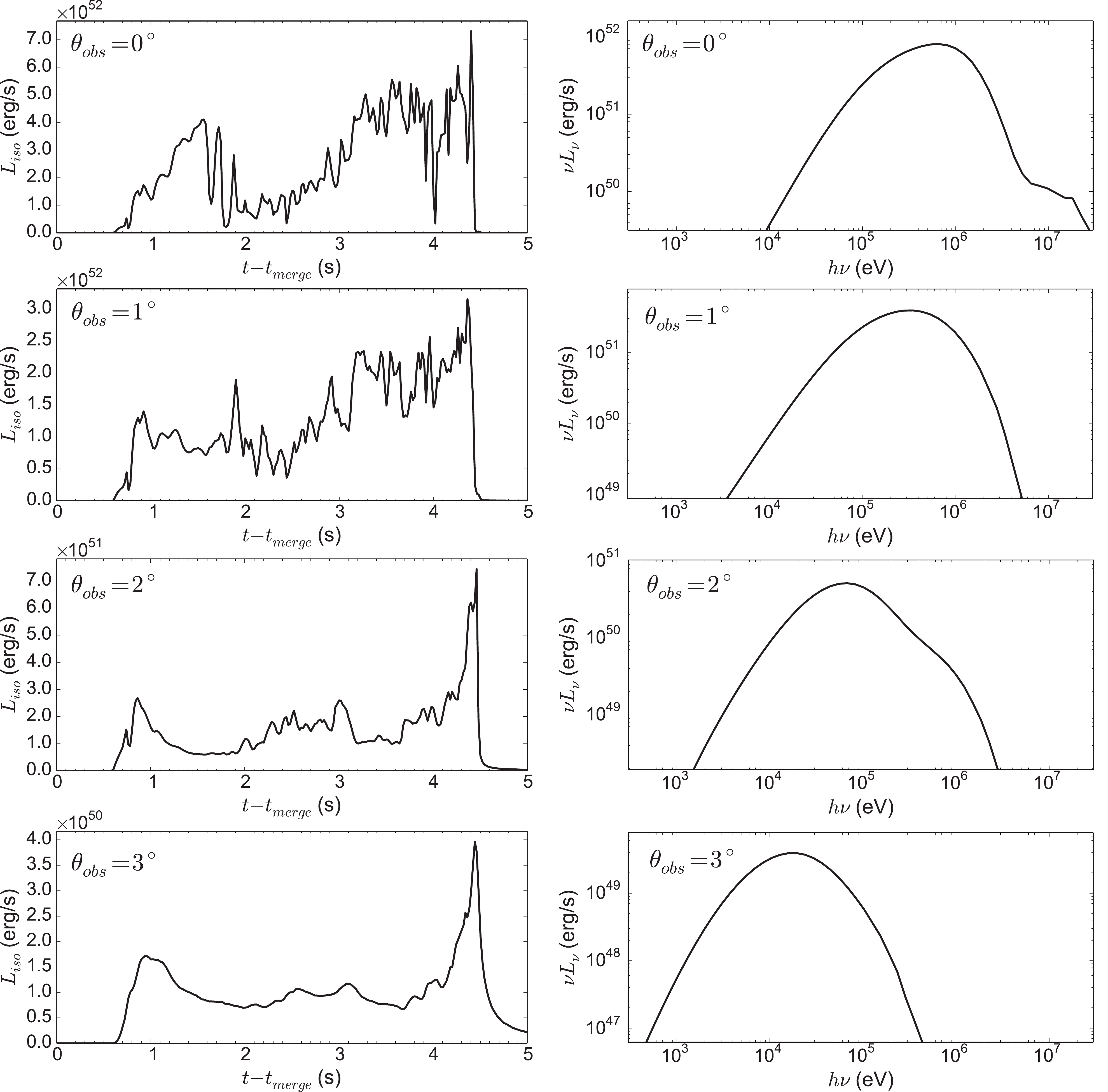}
\end{center}
\caption{Isotropic equivalent luminosity ({\it left column}) and time averaged spectra ({\it right column}) for viewing angles $\theta_{\rm obs} = 
0^{\circ}, 1^{\circ}, 2^{\circ}$ and $3^{\circ}$ ({\it from top to bottom}). The light curves are computed using discretized time bins of $20\,$ms 
width. The time $t=t_{\rm merge}$ corresponds to the time of the merger in the observer frame, i.e. the moment when the peak of the gravitational wave signal from the BNS coalescence reaches the observer. Note that the delay in the onset of prompt emission after $t=t_{\rm merge}$ is dominated by the lifetime of the metastable neutron-star remnant, which we assume to be $\tau_{\mathrm{MNS}}=500$\,ms in our model. The spectra are time averaged between $t-t_{\rm merge} = 0$ and $5\,$s.}
\label{fig:4}
\end{figure*}

The light curves and time-integrated spectra resulting from the relativistic outflow are displayed in Fig.~\ref{fig:4} for various viewing angles. 
The time variability in the light curves traces the structure of the jet, 
which itself is to a large degree shaped by the jet-torus interaction as discussed in Sect.~\ref{sec:hydr-simul}. We observe rapid variability in the light curves down to timescales of $\sim 10-100~{\rm ms}$. It should be noted, however, that these values roughly coincide with the lower limit of numerically resolvable structures with our radial grid and, hence, a 
higher resolution may reveal even shorter variability timescales. As can be seen in Fig.~\ref{fig:4}, the variability timescale tends to become longer at larger viewing angles, because there the photospheric radius, $r_{\rm ph}$, is larger and the corresponding Lorentz factor, $\Gamma_{\rm ph}$, is smaller, which effectively increases the lower bound for the variability timescale  $\sim r_{\rm ph}/2\Gamma_{\rm ph}^2 c$ \citep{P04}. 
While $r_{\rm ph}/2\Gamma_{\rm ph}^2 c < 10~{\rm ms}$ is satisfied at $\theta_{\rm obs} \lesssim \theta_{\rm core}$, the minimum timescale increases rapidly at larger viewing angles, $\theta_{\rm obs}$. For instance, a variability timescale of $\sim  50~{\rm ms} - 200~{\rm ms}$ is found at $\theta_{\rm obs} = 3^{\circ}$.

As seen in Fig.~\ref{fig:4}, the spectra of the emitted radiation are broader than the blackbody spectrum. 
The broadening is due to the multi-color temperature effect (i.e. superposition of photons decoupling from various locations with different temperatures)
as well as bulk Comptonization arising from velocity shear in the outflow 
\citep[e.g.,][]{INO13, INM14}.
%
However, while they may be compatible with some fraction of GRBs, our spectra are still somewhat narrower than those found in typical GRBs.
Such a tension with the observations is commonly found for simulation-based models of photospheric emission \citep{IMN15,L16, PL18, PLL18, IMN19}.
This issue may, at least partially, be mitigated by better resolving the sharp shear structures, which tend to be smeared out by the limited spatial resolution \citep{LPR14, INM14,  PLL18}.
Moreover, small-scale dissipative processes not captured by our simulations \citep[e.g.,][]{VB16, ILS18, LB19} may provide additional non-thermal broadening.
%

\subsection{$E_p-L_p$ and $E_p-E_{\rm iso}$ correlations}\label{sec:e_p-l_p-e_p}

For viewing angles that lie within the opening angle of the core ($\theta_{\rm obs} \lesssim  1.2^{\circ}$), the spectral peak energy, $E_p$, does 
not vary strongly with viewing angle and is about a few times $100~{\rm keV}$, i.e. consistent with typical observed GRBs. Likewise, also the photon luminosity remains fairly constant within $\theta_{\rm obs}<\theta_{\rm core}$.
Going towards larger viewing angles, the peak energy and the luminosity show a rapid decline, reflecting the sharp gradient in the wing region of the jet. Hence, as also found for simulations of LGRBs \citep{PLL18, IMN19}, $E_p$ correlates with the luminosity and with the energy of the radiation as a result of the viewing angle dependence.

\begin{figure*}[htbp]
\begin{center}
  \includegraphics[width=17.5cm,keepaspectratio]{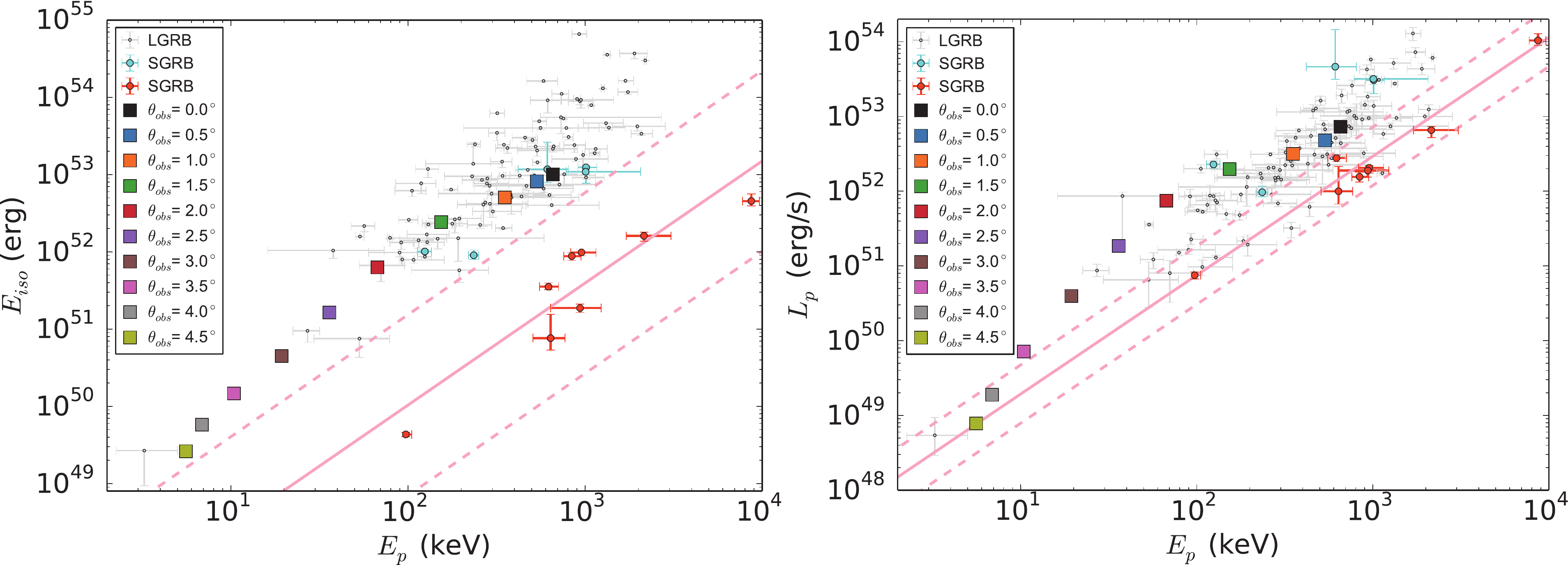}
\end{center}
\caption{Isotropic equivalent energy, $E_{\rm iso}$ ({\it left}), and peak luminosity, $L_p$ ({\it right}), versus the peak energy, $E_p$, of the prompt emission resulting in our model for different observation angles ({\it colored squares}).
  For comparison, observational data of LGRBs taken from \citet{YMT10} ({\it gray points}) and SGRBs taken from \citet{TYN13} ({\it cyan and red circles}) was added. According to the analysis of \citet{TYN13}, red circles represent secure SGRBs, while cyan circles are considered as misguided 
SGRBs. The best-fitting function and $3\sigma$ intervals of the Amati ($E_p - E_{\rm iso}$) and Yonetoku ($E_p - L_p$) relations for the secure SGRBs are indicated with red solid and dashed lines, respectively.
  %
}
\label{fig:5}
\end{figure*}

The relation between $E_p$ and the peak luminosity, $L_p$, of the light curve, as well as between $E_p$ and the total energy of the emission, $E_{\rm iso}$, are summarized in Figure \ref{fig:5}, together with the samples of observed LGRBs and SGRBs that were analyzed in \citet{YMT10} and \citet{TYN13}, respectively.
In \citet{TYN13}, SGRBs are classified into two categories: ``secure'' SGRBs (red circles), which are likely to stem from a different physical origin than typical LGRBs, and ``misguided'' SGRBs (cyan) which may have a similar origin as typical LGRBs.
Similar to previously investigated models of LGRBs \citep{PLL18,IMN19}, 
the slopes of the $E_p-E_{\rm iso}$ and $E_p - L_p$ correlations found for the numerical models show a very good agreement with those of the observed Amati- and Yonetoku-relations, respectively \citep{AFT02, YMN04}.
However, while the distribution from our model overlaps with the population of LGRBs and misguided SGRBs, it lays outside of (i.e. above) the $3-\sigma$ region of secure SGRBs.
The reason for this discrepancy, which points to shortcomings of our model, or possibly just unfavorably chosen input parameters, remains unclear and needs to be explored in upcoming work.

\subsection{Properties of polarization}\label{sec:prop-polar}

\begin{figure}[htbp]
\begin{center}
  \includegraphics[width=9cm,keepaspectratio]{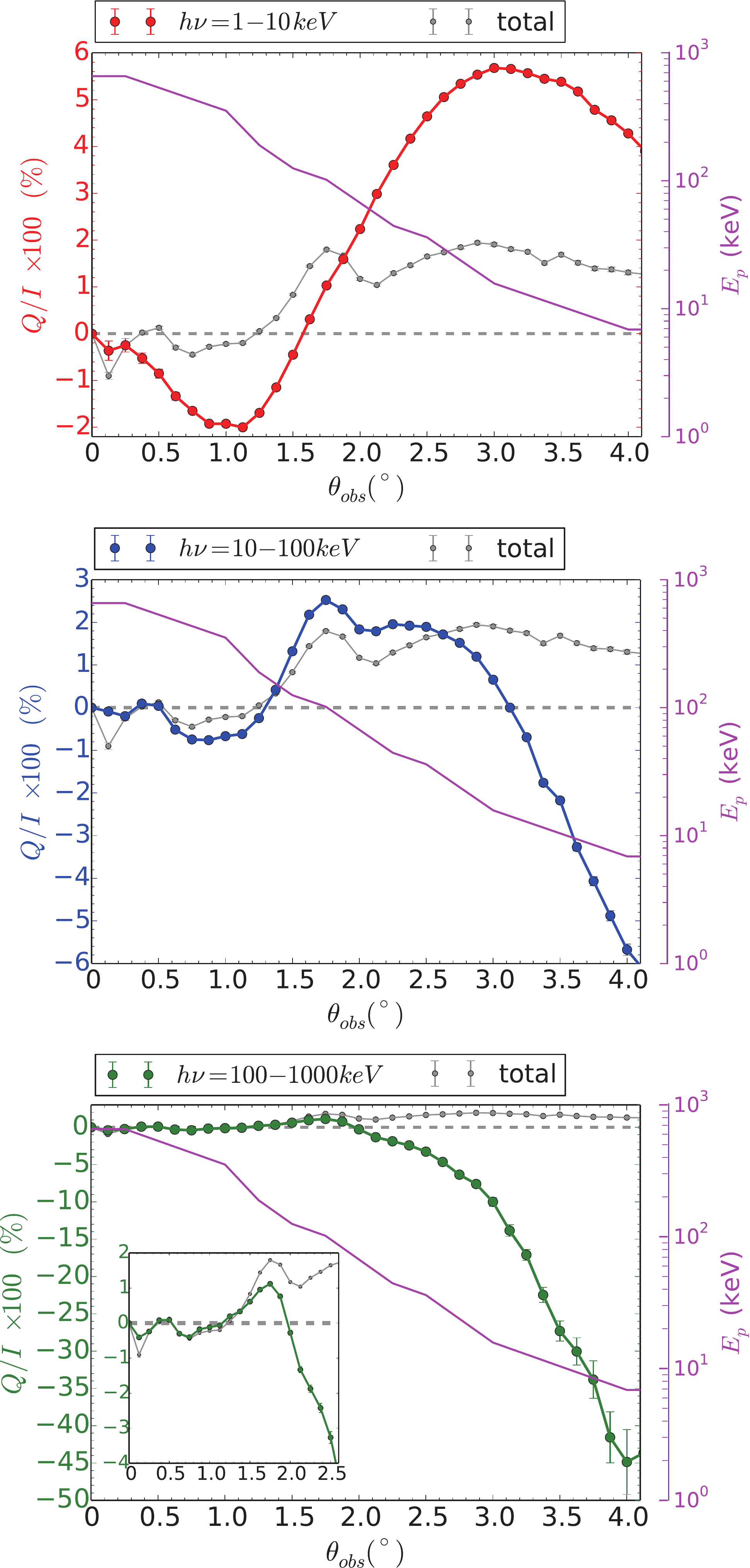}  
\end{center}
\caption{Linear polarization ratio ($\Pi = Q/I$) as function of viewing 
angle at energy ranges of $h\nu = 1 - 10~{\rm keV}$ ({\it top}), $10-100~{\rm keV}$ ({\it middle}) and  $100-1000~{\rm keV}$ ({\it bottom}), as well as the energy integrated (total) polarization ({\it grey solid lines}), the location of $\Pi=0$ ({\it dashed lines}), and the peak energy, $E_p$ ({\it magenta lines}). Error bars indicate the $1\sigma$ statistical uncertainty estimated assuming error propagation as
  $\sigma^2 = \left(\frac{\partial \Pi}{\partial I}\right)^2 \sigma_I^2 
+ \left(\frac{\partial \Pi}{\partial Q}\right)^2 \sigma_Q^2 + 2 \left(\frac{\partial \Pi}{\partial I}\right) \left(\frac{\partial \Pi}{\partial Q}\right) \sigma_{IQ}$, where $\sigma_{I/Q}^2$ and   $\sigma_{IQ}$  denote the variance and covariance, respectively, of the Stokes parameters $I$ and $Q$. Note that $\Pi$ is constrained to vanish in axisymmetry at $\theta_{\rm obs}=0^{\circ}$. The inset in the bottom panel provides an enlarged view of the data for $\theta_{\rm obs} \leq 2.5^{\circ}$.}
\label{fig:6}
\end{figure}

The results of the polarization analysis are displayed in Fig.~\ref{fig:6}.
As seen in the figure, the degree of polarization, $|Q|/I$, observed outside of the jet core 
(i.e. for $\theta_{\rm obs} \gtrsim  1.2^{\circ}$) tends to be  larger than that within
the core ($\theta_{\rm obs} \lesssim 1.2^{\circ}$).
This is reflecting the sharp viewing angle dependence of the emission in the wing region ($\theta > \theta_{\rm core}$), which results in an enhanced level of anisotropy of emission from around the LOS.
This tendency is consistent with analytical studies \citep{LPR14, INM14}.
It should be noted, however, that, while these analytical studies find an 
energy integrated (i.e. total) polarization as high as $|Q|/I \sim 30-40\,\%$, we find much smaller values for this quantity ($\lesssim 2\,\%$: see grey lines in Figure \ref{fig:6}). This can simply be explained by the larger gradient in the lateral direction imposed in the outflow models of 
the analytical studies. Our current result is broadly consistent with the 
LGRB simulations of \citet{PLL20}, in which likewise a low level of polarization was reported.

Similar to what was found in \citet{INM14}, the current simulation also exhibits a significant dependence of the polarization ratio on the energy band.
Compared to the energy integrated polarization, $|Q|/I$ tends to be higher at a given energy band. When viewing into the jet core ($\theta_{\rm obs} \lesssim 1.2^{\circ}$) the low energy part of the spectrum ($h \nu = 
1-10\,\mathrm{keV} \ll E_p$) shows the largest polarization, although it still remains at a rather low level of $\lesssim 2\,\%$. In contrast, at larger viewing angles higher energy bands tend to show larger polarization: While a moderate level of polarization of $\sim 1-6\,\%$ is observed for $h\nu \leq 100~{\rm keV}$, at higher energies of $h\nu = 10^2-10^3~{\rm keV}$ the polarization ratio becomes as high as $\sim 10-40\,\%$ at $\theta_{\rm obs} \gtrsim 3^{\circ}$. The large polarization is attributed 
to the rapid decline of the luminosity with higher values of $\theta_{\rm 
obs}$
in the high energy domain ($h\nu \gg E_p$) of the spectrum.
It is worth to note that not only the magnitude but also the direction of 
the polarization (i.e. the sign of $Q$) varies across the energy bands.

Regarding implications for observations, our model suggests that typical GRBs (which have $E_p \gtrsim 100~{\rm keV}$) should exhibit a low degree 
of polarization (from $\sim 0\,\%$ up to $\sim 2\,\%$ at most). This is consistent with the latest report by the POLAR mission \citep{ZKB19,KAB20}, which finds that the emission in the sample of 14 GRBs, of which 2 are classified as SGRBs, is consistent with having a low polarization level or even being unpolarized. On the other hand, the possibility inferred in their analysis that these low levels of polarization might be due to a changing polarization angle, which effectively washes out the polarization, 
is not compatible with our model. 
While our time-resolved analysis also reveals a temporal variation of the 
sign of $Q$ (which corresponds to the change in the polarization angle by 
$90^{\circ}$), the instantaneous degree of polarization at a given time is much weaker ($\lesssim $ few$~\%$) than the values suggested by the aforementioned studies for $\theta_{\rm obs} \lesssim \theta_{\rm core}$. Moreover, our model is inconsistent with earlier observational studies in which quite high polarization levels are inferred \citep[e.g., by the GAP instruments,][]{YMG11, YMG12}. However, since all measurements up to date suffer from low photon statistics these comparisons are not conclusive.

Our result also predicts that dim bursts with soft spectra ($E_p \lesssim 
100~{\rm keV}$) tend to show stronger polarization. In particular, at $h\nu \gtrsim E_p$ a significant polarization ratio of $\gtrsim 10~\%$ may be encountered in the emission. While being quite challenging to test with 
observations, this trend could in principle be used to probe the emission 
mechanism of GRBs.

\section{Comparison with GRB170817A/GW170817}\label{sec:1701817}

In the previous section, we have focused on the emission from small viewing angles ($\theta_{\rm obs} \lesssim 4 \theta_{\rm core}$) in order 
to compare our results to ordinary observed SGRBs. We now compare our results to GRB170817A. This event was a peculiar GRB, which was orders of magnitude fainter than the faintest SGRB observed thus far, while it exhibited a spectral peak energy as high as in typical bursts \citep[$E_{\rm iso} = (4.8 \pm 0.8)\times 10^{46}\,$erg, $L_p = (1.4 \pm 0.5) \times 10^{47}$\,erg\,s$^{-1}$, and $E_p = 185 \pm 62\,$keV; see][]{GVB17, VMG18}. According to the observational constraints from the combination of  VLBI 
 and afterglow modeling, the viewing angle of this burst is estimated to lie in the range $\theta_{\rm obs} \sim 14^{\circ} - 19^{\circ}$ \citep{MDG18, GSP19, HNG19}.
While our radiative transfer simulation does not cover such a large viewing angle, the sharp drop of $E_p, E_{\rm iso}$, and $L_p$ with the viewing angle beyond the core of the jet (cf. Fig.~\ref{fig:5}) strongly suggests that the photospheric emission component of our model would have difficulties to reproduce the observed properties of GRB170817A.
%
Thus, it is quite likely that a different emission process was playing the dominant role in the burst.

One possibility is that the gamma-rays are produced by some internal dissipation processes
that operate in the far off-axis region and are not captured by our model \citep[e.g.,][]{KBG18, IN19}. However, their relevance and efficiency are poorly understood, which is why no robust assessment of this possibility can be made at this point.

An alternative and physically well motivated scenario is that the emission is produced by the shock breakout of the cocoon, which inevitably accompanies the jet breakout \citep[for a comprehensive overview, see][]{N20}. 
While its properties depend sensitively on the detailed conditions of the 
shock breakout (e.g. density and velocity profile of the ejecta and the strength of the shock propagating through the ejecta), its luminosity is generically much lower than that of typical GRBs. This is due to the fact that the breakout emission releases only the energy stored in a small fraction of the ejecta that is located at its outermost edge. However, since 
the cocoon shock breakout releases radiation into a larger solid angle, it is likely to dominate the narrowly collimated jet emission at high latitudes. 
Previous studies have indeed shown that within a reasonable setup, the shock breakout signal
is capable of reproducing the observed properties of GRB170817A \citep{KNS17, GNP18, NGP18, PBM18, BLL20, LB21}.

The shock breakout scenario does not contradict with the results of our current model.
In fact, the sharp decline of the luminosity outside of the core region (cf. Fig.~\ref{fig:5}) indicates that the photospheric component is likely 
much weaker than the shock breakout emission at large viewing angles.
Note that, although a broad cocoon component is indeed present in our hydrodynamical model, our computational methods lack the ability to resolve the breakout emission.\footnote{The current radiative transfer calculation 
is optimized to capture the diffusion of photons that are originally produced within the jet deep below the photosphere, and therefore it cannot resolve photons produced during the shock breakout \citep[][]{ILN20, ILN20b}.}
Therefore, we cannot give a quantitative estimate of the angle at which the breakout emission would start to dominate the photospheric emission. Nonetheless, assuming that the cocoon breakout emission produces a signal comparable to that of GRB170817A, it is quite likely that the cocoon emission is dominant at large viewing angles such as in GRB170817A.

To sum up, we speculate that the emission outside of the jet core is dominated by a photospheric component from the moderately relativistic wings up to a certain threshold angle, above which the shock breakout component 
dominates. As discussed in the previous section, below this threshold angle the generic decline of $E_p, L_p$,~and~$E_{\rm iso}$ with growing viewing angle $\theta_{\rm obs}$ is expected to produce the $E_p - L_p$ and $E_p - E_{\rm iso}$ correlations observed in ordinary SGRBs \citep{TYN13}. 
On the other hand, above this threshold angle, where the cocoon shock breakout emission dominates, the angular dependence of the emission should become much weaker. Hence, emission from the cocoon shock breakout does not follow the $E_p - L_p$ ($E_{\rm iso}$) correlations, and it is expected 
to appear as an outlier of low luminosity (energy) and high $E_p$ in the $E_p - L_p$ ($E_{\rm iso}$) diagram, as is the case for GRB170817A.
It is worth to note that 
  dimmer and softer GRBs that originate from the wing region (i.e. $\theta_{\rm obs} \geq \theta_{\rm core}$) have a much larger chance of being detected as the electromagnetic counterpart to a gravitational wave event than 
face-on GRBs (where $\theta_{\rm obs} \leq \theta_{\rm core}$) owing to the larger solid angle.\footnote{If for our current model we assume, for instance, that the emission from the wing is dominant at least up to $\sim 
4.5^{\circ}$, the solid angle of the wing would be more than a factor of 10 greater than the solid angle of the core ($\theta_{\rm core} \approx 1.2^{\circ}$).}

\section{Summary and discussion}\label{sec:summary}

In this study, we have computed the prompt photospheric emission from a jet that breaks out of BNS merger ejecta by post-processing the results of a hydrodynamic simulation with a gamma-ray radiative transfer solver. To our knowledge, this is the first such study conducted in the context of SGRBs. The jet was injected ``by hand'' with a constant power of $10^{50}\,$erg\,s$^{-1}$ and opening angle of $10^\circ$, but in contrast to most previous jet studies, we consistently followed the evolution of the BH-torus system and its interaction with the jet, while taking into account neutrino transport and viscous angular momentum transport. We analyzed the light curves, spectra, and polarization properties of the photospheric emission radiated from the relativistic outflow.

Our main findings are as follows:


\begin{enumerate}
\item The jet is strongly affected by the presence of the disk. The wind expelled by the disk acts as a collimating agent that regulates the jet opening angle already before the jet enters the ejecta originating from the metastable NS and the dynamical ejecta. Near the jet base ($r\la 10^9\,$cm) we approximately observe smaller (larger) opening angles for higher (lower) disk wind mass fluxes.
  Hence, the role of the disk wind for determining the final jet opening angle may be comparable to, and may for some BNS configurations even dominate, the role of the other ejecta components that has been widely studied in previous works \citep[e.g.,][]{NHS14, MMR14,  DQM15,  MRM17, LLC17, DQK18, GNP18, GLN19,  HKI20, HI21, GNB21}.

\item The final jet opening angle in our model ($\theta_{\rm core} \sim 1.2^{\circ}$) turns out to be smaller than those inferred from observations of typical SGRBs \citep[$\gtrsim 3^{\circ}$;][]{FBM15} as well as of GRB170817A \citep[$\sim 2^{\circ}-4^{\circ}$;][]{MDG18, GSP19, TER19}. Moreover, in contrast to previous studies neglecting the central torus evolution (and mostly considering LGRB jets), who report a ratio of the final 
core opening angle to the opening angle of the initially injected jet of $\theta_{\rm core}/\tilde\theta_{j, \rm ini} 
\sim \frac{1}{5}\ldots \frac{1}{3}$ \citep[][]{Mizuta2013a, GLN19, GNB21}, we find $\theta_{\rm core}/\tilde\theta_{j, \rm ini} \sim \frac{1}{10}$ for our current model  (where $\tilde\theta_{j,\rm ini}\equiv \theta_{j, \rm ini} + 0.7\Gamma_{j,\rm ini}^{-1}$ is the effective initial opening angle). The sensitivity of $\theta_{\rm core}$ to variations of the model parameters and the reason for the relatively narrow GRB jet resulting in our model need to be further explored in future work.

\item The jet-torus interaction not only enhances the jet collimation but 
it also imprints rapid variability on the jet down to timescales of $\sim 
10~{\rm ms}$. This variability is generated by perturbations traveling along the disk wind, which ultimately have their origin in the turbulent disk. Consistent with the variability of the jet, we observe temporal variations of $\sim 10-100\,$ms in the light curve emanating from the jet core 
(i.e. for $\theta_{\rm obs} \lesssim \theta_{\rm core} \approx 1.2^{\circ}$). Interestingly, this is consistent with the variability timescales of 
observed SGRBs \citep[e.g.,][finds a median value of $\sim 10~{\rm ms}$ for the minimum variability timescale]{GBL15}. The variability timescale becomes longer for larger viewing angles. This tendency suggests that a dimmer soft burst should exhibit a smoother light curve, and it may therefore, at least partially, explain the negative correlation between the luminosity and minimum variability timescale inferred for observed GRBs \citep{SMD15}.
  
\item Similar to simulations of LGRBs \citep{PLL18,IMN19}, we find that the viewing angle dependence of the emission leads to correlations between 
$E_p$ and $L_p$ as well as $E_p$ and $E_{\rm iso}$ with a slope broadly consistent with observations. This finding suggests that these correlations are inherent features of photospheric emission also in the context of SGRBs. However, while the slope of these correlations is consistent with observed SGRBs, our model exhibits a notable offset. The reason for this discrepancy is unclear but it may be connected to the same circumstances that lead to the small opening angle of our jet, $\theta_{\rm core}$.

\item The fact that our model cannot reproduce GRB1701817A, which is a clear outlier in the correlation diagram, supports the scenario that this event was no ordinary SGRB but was likely produced by a cocoon shock breakout (\citealp[e.g.][]{N20}).

\item The degree of polarization, $|Q|/I$, strongly depends on the viewing angle and energy band. It is low ($\lesssim 2\,\%$) for small viewing angles $\theta_{\rm obs} \lesssim \theta_{\rm core}$ (where $E_p \gtrsim 100~{\rm keV}$), which represents the typical case for observed SGRBs. On the other hand, for soft dim bursts ($E_p \lesssim 100~{\rm keV}$), which 
are observed at larger viewing angles, the degree of polarization tends to increase with $\theta_{\rm obs}$.
We find significant polarization particularly for the high energy portion 
($h\nu \geq E_p$) of the emission, where it can become as high as $\sim 10 - 40\%$ at large viewing angles $\theta_{\rm obs} \gtrsim 2.5\times\theta_{\rm core}$.
Apart from a varying degree of polarization, we also observe that the angle of the polarization (i.e. the sign of $Q/I$) can differ among energy bands.
These characteristics could in principle be used to probe the nature of the prompt emission mechanism.
The current result is consistent with the latest observation by POLAR  \citep{ZKB19,KAB20}, which suggests low or unpolarized emission in bright GRBs. 
However, all observations up to now suffer from large uncertainties, and more precise measurements will be required to formulate robust constraints. Future missions, such as POLAR-2 \citep{HAB21} and  LEAP \citep{WMA21}, may be promising in that respect.

\end{enumerate}

Before concluding, we comment on the limitations of our model. As already mentioned in Sect.~\ref{sec:light-curves-spectra}, the limited grid resolution, which hardly captures all relevant details of the shear flow structures around the photosphere, may, at least partially, explain why our spectra are narrower than those of typical GRBs \citep{LPR14, INM14, PLL18}. 
Moreover, sub-photospheric dissipative processes such as radiation mediated shocks \citep[e.g.,][]{ILS18, LB19} and/or those which invoke generation of relativistic electrons/positrons \citep[e.g.,][]{VB16} 
are not taken into account.
%
Since the aforementioned effects are likely to broaden the spectrum, the current result should be considered as a lower bound regarding non-thermal features. It should also be noted that, if dissipation 
gives rise to synchrotron emission, this may also affect the polarization 
properties \citep{LVB18}.
An additional shortcoming of our model is the restriction to 2D axisymmetry, which is known to artificially suppress mixing in the jet-cocoon interface by inhibiting non-axisymmetric instability modes \citep{Harrison2018o,GNB21, PCV21} and which can overestimate the amount of mass accumulated in front of the jet \citep{Zhang2003, Harrison2018o}.
Finally, our model is purely hydrodynamical. Weakly magnetized jets were found \citep{GBS20} to be less susceptible to instabilities of the jet-cocoon interface (which may, however, effectively reduce again the aforementioned discrepancy between 2D and 3D models). Moreover, in the case of a general relativistic MHD evolution the jet is launched self-consistently due to the Blandford-Znajek process and does not need to be parameterized as in our case \citep[see, e.g.,][for MHD models of SGRB jets]{Kiuchi2015, FTQ19, KTG19, CLT19, NGP21}. However, 3D MHD models are not only much more expensive than our 2D model, they also introduce additional uncertainties and difficulties, e.g. connected to numerical convergence \citep[see, e.g.,][]{Liska2020a} or to the choice of the initial magnetic field distribution \citep{CLT19}.


\acknowledgments
 We thank T. Parsotan, M. Barkov, A. Mizuta and D. Warren for fruitful discussions. This work was supported by JSPS KAKENHI Grant Number JP19K03878, JP19H00693, JP20H04751 and JP17H06362. OJ was supported by the Special 
Postdoctoral Researchers (SPDR) program at RIKEN and by the European Research Council (ERC) under the European Union's Horizon 2020 research and innovation programme under grant agreement No. 759253. YT was supported by 
the RIKEN Junior Research Associate Program. Numerical computations and data analysis were carried out on Hokusai BigWaterfall system at RIKEN,   XC50  at Center for Computational Astrophysics, National Astronomical Observatory of Japan and the Yukawa Institute Computer Facility. This work was supported in part by a RIKEN Interdisciplinary Theoretical \& Mathematical Science Program (iTHEMS) and a RIKEN pioneering project ``Evolution of Matter in the Universe (r-EMU)'' and ``Extreme precisions to Explore fundamental physics with Exotic particles (E3-Project)''.


\end{document}